\documentclass[twocolumn,showpacs,preprintnumbers,amsmath,amssymb]{revtex4}

\usepackage{graphicx}
\usepackage{dcolumn}
\usepackage{bm}

\newcommand{\kagome}{ {kagom\'e} }

\begin{document}

\title{Low-energy singlet dynamics of spin-$\frac12$ \kagome Heisenberg antiferromagnets}

\author{A. V. Syromyatnikov}
 \email{syromyat@thd.pnpi.spb.ru}
\author{S. V. Maleyev}
\affiliation{Petersburg Nuclear Physics Institute, Gatchina, St.\ Petersburg 188300, Russia}

\date{\today}

\begin{abstract}

We give a physical picture of the low-energy sector of the spin-$\frac12$ \kagome Heisenberg antiferromagnets (KAFs). It is shown that \kagome lattice can be represented as a set of blocks containing 12 spins, having form of stars and arranged in a triangular lattice. Each of these stars has two degenerate singlet ground states which can be considered in terms of pseudospin-$\frac12$. It is shown using symmetry consideration that the KAF lower singlet band is made by inter-star interaction from these degenerate states. We demonstrate that this band is described by effective Hamiltonian of a magnet in the external magnetic field. The general form of this Hamiltonian is established. The Hamiltonian parameters are calculated up to the third order of perturbation theory. The ground state energy calculated in the considered model is lower than that evaluated numerically in the previous finite clusters studies. A way of experimental verification of this picture using neutron scattering is discussed.  It is shown that the approach presented can not be expanded directly on KAFs with larger spin values.

\end{abstract}


\maketitle

\section{Introduction}

Unusual low-temperature properties of \kagome antiferromagnets (KAFs) attracted much attention of both theorists and experimenters in last decade. Seemingly the most intriguing features were observed in specific heat $C$ measurements of SrCr$_{9p}$Ga$_{12-9p}$O$_{19}$ (spin-$\frac32$ \kagome material) \cite{ramirez}. It has been revealed a peak at $T\approx 5$ K which was practically independent of magnetic field up to 12 T and $C$ appeared to be proportional to $T^2$ at $T\alt 5$ K.

There was no appropriate theory describing KAF low-energy sector so far. Qualitative understanding of the KAF low-temperature physics was based mostly on results of numerous finite cluster investigations \cite{lecheminant,zeng,leung,wald,zeng2}. They revealed a gap separating the ground state from the upper triplet levels and a band of nonmagnetic singlet excitations with a very small or zero gap inside the spin gap. The number of states in the band increases with the number of sites $N$ as $\alpha^N$. It was obtained for samples with up to 36 sites that $\alpha=1.15$ and 1.18 for even and odd $N$, respectively \cite{lecheminant,wald}. It is believed now that this
wealth of singlets is responsible for specific heat low-$T$ peak and explains its field independence \cite{sind,ramirez}.

The origin of the singlet band as well as the nature of the ground state were unclear until now. Previous exact diagonalization studies of clusters with $N\le36$ \cite{chalker,leung} revealed exponential decay of the spin-spin and dimer-dimer correlation functions. So the point of view that KAF is a spin liquid is widely accepted \cite{sind,lecheminant,sachdev,zeng2,yang,kalmeyer,marston,wald,leung,chalker}.

It seems the best candidate for description of KAF low-energy properties is a quantum dimer model \cite{sachdev,zeng2,mila1}. In Ref.~\cite{mila2} an approach firstly pioneered by Subrahmanyam in Ref.~\cite{subr} was developed in which a spin-$\frac12$ \kagome lattice is considered as a set of interactive triangles with a spin in each apex. It was suggested there to work in the subspace where the total spin of each triangle is 1/2 investigating low-lying excitations. In spite of certain success of this approach (coincidence of the low-energy spectrum and the number of lower singlet excitations in samples with up to 36 sites with exact diagonalization results) a further development of this approach is required to get the full physical description of KAF.

Recently in Ref.~\cite{zhit} it was suggested a model of frustrated antiferromagnet which low-energy properties can be generic for KAF as well as for two other types of frustrated magnets which possess a similar behavior as KAF and have many singlet states inside the triplet gap: pyrochlore \cite{canals} and $\rm CaV_4O_9$ \cite{alb}. Weakly interactive plaquettes on the square lattice were considered there. Each plaquette has two almost degenerate singlet ground states, so a band of singlet excitations arises if the inter-plaquette interaction is taken into account. It is shown that there is a quantum phase transition in the model at a critical value of frustration separating a disorder plaquette phase and a columnar dimer one. In the proximity of this transition the specific heat has a low-$T$ peak below which it possesses a power low-temperature dependence.

In our recent paper Ref.~\cite{syromyat} we have shown that a similar picture is relevant for spin-$\frac12$ KAF. It was proposed to consider a \kagome lattice as a set of stars with 12 spins arranged in a triangular lattice (see Fig.~\ref{lattice}). Numerical diagonalization has shown that star has two degenerate singlet ground states separated
from the upper triplet levels by a gap. These states form a singlet energy band as a result of inter-star interaction. It was assumed that this band determines the KAF low-energy singlet sector. We have shown that it is described by Hamiltonian of a magnet in the external magnetic field where degenerate states of the stars are represented in terms of two projections of pseudospin-$\frac12$.

The main advantage of a star over a triangle proposed in Refs.~\cite{mila2,subr} as a starting point to study KAF is that consideration of star two singlet ground states explicitly leads to effective Hamiltonian which describes singlet low-temperature dynamics.

In the present more comprehensive paper we develop this approach. It is proved using symmetry considerations presented in Sec.~\ref{sym_con} that the singlet band above discussed determines the KAF lower singlet sector.

This band is studied in Sec.~\ref{sin_dyn}. The general form of the effective Hamiltonian is established there. The Hamiltonian parameters are calculated up to the third order of perturbation theory. The ground state energy calculated in the considered model is lower than that evaluated numerically in the previous finite clusters studies. We demonstrate that our approach can not be expanded directly on KAFs with larger spin values. A way of experimental verification of this picture using neutron scattering is discussed.

Finally we summarize our results in Sec.~\ref{con}.

\section{Symmetry consideration}
\label{sym_con}

\begin{figure}
  \centering
  \includegraphics{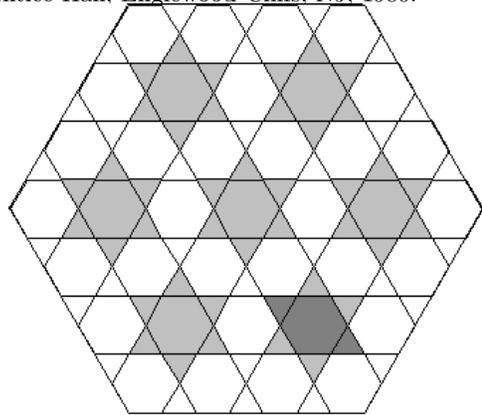}
\caption
{
Kagom\'e lattice (KL). There is a spin in each lattice site. The KL can be considered as a set of stars arranged in triangular lattice. Each star contains 12 spins. A unit cell is also presented (dark region). There are four
unit cells per star.
\label{lattice}
}
\end{figure}

We start with the Hamiltonian of the spin-$\frac12$ \kagome Heisenberg antiferromagnet:
\begin{equation}
\label{h}
{\cal H}_0 = J_1\sum_{\langle i,j \rangle}{\bf S}_i{\bf S}_j + J_2\sum_{(i,j)}{\bf S}_i{\bf S}_j,
\end{equation}
where $\langle i,j \rangle$ and $(i,j)$ denote nearest and next-nearest neighbors on the \kagome lattice, respectively, shown in Fig.~\ref{lattice}.  The case of $|J_2|\ll J_1$ is considered in this paper. We discuss a possibility of both signs of next-nearest-neighbor interactions --- a ferromagnetic and an antiferromagnetic one. As is shown below, in spite of the smallness the second term in Eq.~(\ref{h}) can be of importance for the low-energy properties.

Kagom\'e lattice can be represented as a set of stars arranged in a triangular lattice (see Fig.~\ref{lattice}). To begin with we neglect interaction between stars and put $J_2=0$ in Eq.~(\ref{h}). A star is a system of 12 spins. Let us consider its properties in detail.

As the Hamiltonian Eq.~(\ref{h}) commutates with all the projections of the total spin operator, all the star levels are classified by the values $S$, irreducible representations (IRs) of its symmetry group and are degenerated with $S_z$. The star symmetry group $C_{6v}$ contains six rotations and reflections with respect to six lines passing through the center. There are four one-dimensional and two two-dimensional IRs which are presented in Appendix~\ref{ir}. In their basis the matrix of the Hamiltonian has a block structure. Each block has been diagonalized numerically.

As a result it was obtained that the star has doubly degenerated singlet ground state separated from the lower triplet level by a gap $\Delta\approx0.26J_1$. Ground state wave functions can be represented as follows:
\begin{eqnarray}
\Psi_+&=&\frac{1}{\sqrt{2+1/16}}(\phi_1+\phi_2), \label{f1}\\
\Psi_-&=&\frac{1}{\sqrt{2-1/16}}(\phi_1-\phi_2), \label{f2}
\end{eqnarray}
where functions $\phi_1$ and $\phi_2$ are shown schematically in Fig.~\ref{grounds}. A bold line there represents the singlet state of the corresponding two spins, i.e. $({\mid\uparrow\rangle}_i{\mid\downarrow\rangle}_j -
{\mid\downarrow\rangle}_i{\mid\uparrow\rangle}_j)/\sqrt{2}$.

It can be shown that $\phi_1$ and $\phi_2$ are not orthogonal: their scalar product is $(\phi_1\phi_2)=1/32$. They contain six singlets each having the energy $-S(S+1)J_1=-3/4J_1$. One can obtain that interaction between singlets does not contribute to the energy of the ground states which is consequently equal to $-4.5J_1$.

Functions $\phi_1$ and $\phi_2$ are invariant with respect to rotations of the star and they transform into each other under reflections. Hence $\Psi_+$ is invariant under all the symmetry group transformations. In contrast, function $\Psi_-$ is invariant under rotations, changes the sign with reflections and is transformed under the representation (\ref{forminus}). So the ground state has accidental degeneracy. As it is shown in the next section, the next-nearest neighbor interaction, which has the same symmetry as the original Hamiltonian, lifts this degeneracy.

KAF containing $\cal N$ non-interactive stars has an energy spectrum with great levels degeneracy when ${\cal N}\gg1$. For example, the ground state degeneracy is $2^{\cal N}$ and that of the lowest triplet level is $3{\cal N}2^{({\cal N}-1)}$. Interaction between stars gives rise to an energy band from every such group of levels and it is a very
difficult task to follow their evolution. Meanwhile group theory allows one to make some conclusions about the KAF low-energy sector. We show now that the singlet band stemmed from the ground state can not be overlapped by those originated from the upper singlet levels.

\begin{figure}
  \centering
  \includegraphics{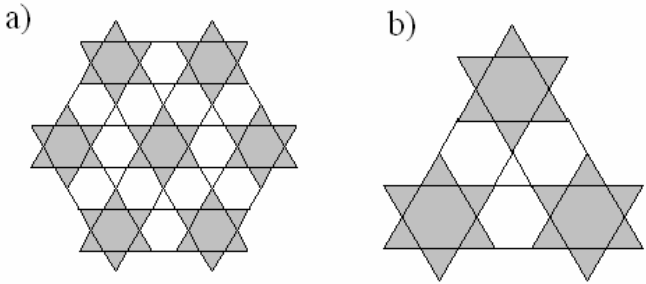}
  \caption
{
\label{clusters}
a) A cluster where operator of interaction between stars has the same symmetry group $C_{6v}$ as the whole cluster; b) The only configuration of three stars giving nonzero contribution to the third term in the perturbation expansion.
}
\end{figure}

Let us consider a cluster with seven stars shown in the Fig.~\ref{clusters}(a) neglecting for a beginning interaction between them.  The symmetry group of the cluster is $C_{6v}$ also. The ground state has degeneracy $2^7=128$. Corresponding wave functions transformed under IRs of $C_{6v}$ are constructed as linear combinations of products containing $\Psi_+$ or $\Psi_-$ for each star. It is easy to show using a standard procedure of irreducible representations basis construction \cite{landau,petrashen} and Appendix~\ref{ir} that there is at least two ground state wave functions of the cluster discussed transformed under any given IR.

It is important to mention that operator of inter-star interaction in the cluster has the same symmetry as the intra-star one which is a sum of the stars Hamiltonians. Then the interaction between stars commutates with the square of the total spin operator. So if we begin to increase the inter-star interaction from zero, all the levels would move by energy but their classification can not be changed. Levels can cross with each other as the interaction arises from 0 to $J_1$ but the crossing is forbidden for those of the same symmetry. It is a consequence of the symmetry theorem proved in Refs.~\cite{petrashen,landau}. Hence, one can conclude that the lower singlet sector of the cluster is formed by states which stem from the origin 128 lower levels.

We can lead to the same conclusions considering clusters of the symmetry $C_{6v}$ with the greater number of stars. So we will assume in the following that the KAF low-energy singlet sector is formed by the states originated from those in which each star is in one of the states $\Psi_+$ or $\Psi_-$. Because bands with $S\ne0$ can overlap the singlet one under discussion we have to suppose that the KAF low-energy properties are determined by the lowest states in this
singlet band.

As interaction between stars commutates with the square of the total spin operator, bands of the different $S$ can be studied independently. The KAF lower singlet sector is considered in detail in the next section. Investigation of states with $S\ne0$ is out of the scope of this paper.

\section{Singlet dynamics}
\label{sin_dyn}

In this section we derive the general form of the effective Hamiltonian described the lower singlet sector. Inter-star interaction is considered as a perturbation. Despite of its not being small comparing to the intra-star one there are reasons presented below to use perturbation expansion here.

\begin{figure}
  \centering
  \includegraphics{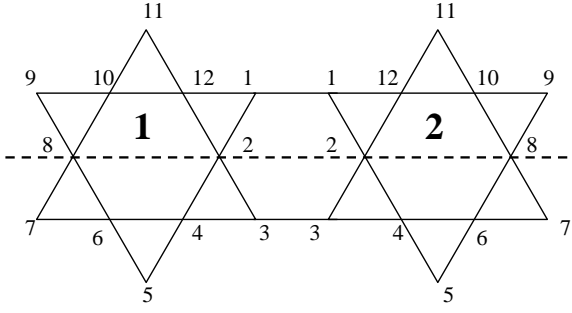}
  \caption
{
Interactions between two stars: $V=J_1({\bf S}^{(1)}_1{\bf S}^{(2)}_1+{\bf S}^{(1)}_3{\bf S}^{(2)}_3)$ and $\tilde V=J_2({\bf S}^{(1)}_1{\bf S}^{(2)}_2+{\bf S}^{(1)}_2{\bf S}^{(2)}_1+{\bf S}^{(1)}_2{\bf S}^{(2)}_3+{\bf S}^{(1)}_3{\bf
S}^{(2)}_2)$, where upper indexes label the stars. The system is symmetric under reflection with respect to the dotted line.
\label{twostars} }
\end{figure}

{\it Two-stars coupling}.---We start with consideration of the interaction between two nearest stars still neglecting the second term in Eq.~(\ref{h}). Initially there is four times degenerated ground state with wave functions $\{
\Psi^{(1)}_{n_1}\Psi^{(2)}_{n_2} \}$ (here $n_i=+,-$ and the upper index labels the stars) and the energy $E_{n_1n_2}^{(0)}=E_{n_1}^{(0)}+E_{n_2}^{(0)}=-9J_1$. As it is seen from Fig.~\ref{twostars}, the interaction has the form
\begin{equation}\label{v}
V=J_1({\bf S}^{(1)}_1{\bf S}^{(2)}_1+{\bf S}^{(1)}_3{\bf S}^{(2)}_3).
\end{equation}
According to the standard theory \cite{landau} the following conditions should be fulfilled to consider $V$ as a perturbation:
\begin{equation}
\label{pertcond}
\left|C^{n_1n_2}_{m_1m_2}\right|=\left|\frac{V_{n_1n_2;m_1m_2}}{E_{n_1n_2}^{(0)}-E_{m_1m_2}^{(0)}}\right|\ll1,
\end{equation}
where $V_{n_1n_2;m_1m_2}=\langle \Psi^{(1)}_{n_1}\Psi^{(2)}_{n_2}\mid V\mid\Psi^{(1)}_{m_1}\Psi^{(2)}_{m_2}\rangle$, $m_1m_2$ denotes excited singlet levels of the two stars and $n_i=+,-$. We have calculated $C^{n_1n_2}_{m_1m_2}$ for $n_i=+,-$ using wave functions obtained numerically and found that all of these coefficients do not exceed $0.09$. So conditions Eq.~(\ref{pertcond}) are fulfilled. Then the maximum value of the sum $\sum_{m_1m_2}|C^{n_1n_2}_{m_1m_2}|^2$ is $0.28$ which is also small enough. Thus interaction between stars will be considered as a perturbation in the following.

We proceed with calculations of corrections to the initial ground state energy of two stars. For this purpose, as the state is fourfold degenerated, one has to solve a secular equation \cite{landau}. The corresponding matrix elements in the third order of perturbation theory are given by \cite{landau}
\begin{multline}
\label{siecle}
		H_{n_1n_2;k_1k_2} = V_{n_1n_2;k_1k_2}+\sum_{m_1,m_2}\frac{V_{n_1n_2;m_1m_2}V_{m_1m_2;k_1k_2}}
    {E_{n_1n_2}^{(0)}-E_{m_1m_2}^{(0)}}\\
    +\sum_{m_1,m_2}\sum_{q_1,q_2}\frac{V_{n_1n_2;m_1m_2}V_{m_1m_2;q_1q_2}V_{q_1q_2;k_1k_2}}
    {(E_{n_1n_2}^{(0)}-E_{m_1m_2}^{(0)})(E_{n_1n_2}^{(0)}-E_{q_1q_2}^{(0)})},
\end{multline}
where  $n_i,k_i=+,-$. Obviously the first term in Eq.~(\ref{siecle}) is zero and the second one can be represented as follows:
\begin{multline}
\label{s2}
    H_{n_1n_2;k_1k_2}=-i\int_0^\infty dt e^{-\delta t+iE_{n_1n_2}^{(0)}t}\\
    \times\langle \Psi^{(1)}_{n_1}\Psi^{(2)}_{n_2}\mid
    Ve^{-it({\cal H}_0^{(1)}+{\cal H}_0^{(2)})}V\mid\Psi^{(1)}_{k_1}\Psi^{(2)}_{k_2}\rangle,
\end{multline}
where ${\cal H}_0^{(i)}$ are Hamiltonians of the corresponding stars. The third term in Eq.~(\ref{siecle}) will be considered later. Using the symmetry of the functions $\Psi_+$ and $\Psi_-$ discussed above and invariance of the system under reflection with respect to the dotted line plotted in Fig.~\ref{twostars}, one can show that only nonzero elements belong to the first and to the second diagonals (i.e.\ with $n_1=k_1$, $n_2=k_2$ and with $n_1\ne k_1$, $n_2\ne k_2$). We have calculated them numerically with a very high precision by expansion of the operator exponent up to the power 150. The results can be represented in the following form:
\begin{subequations}\label{mel}
\begin{eqnarray}
H_{++;++}&=&-a_1+a_2-a_3,\\
H_{+-;+-}&=&-a_1+a_3,\\
H_{-+;-+}&=&-a_1+a_3,\\
H_{--;--}&=&-a_1-a_2-a_3,
\end{eqnarray}
\end{subequations}
where $a_1=0.256J_1$, $a_2=0.015J_1$ and $a_3=0.0017J_1$. Terms of the second diagonal $H_{++;--}=H_{--;++}=-H_{+-;-+}=-H_{-+;+-}=a_4=-0.0002J_1$ are much lesser then $a_1$, $a_2$ and $a_3$. So the interaction shifts all the levels on the value $-a_1$ and lifts their degeneracy. Constants $a_2$, $a_3$ and $a_4$ determine the levels splitting. It is seen that the splitting is very small comparing to the shift.

So KAF appears to be a set of two-levels interacting systems and one can naturally represent the low-energy singlet sector of Hilbert space in terms of pseudospins: ${\mid\uparrow\rangle}=\Psi_-$ and ${\mid\downarrow\rangle}=\Psi_+$. It is seen from Eqs.~(\ref{mel}) that in these terms the interaction between stars is described by the Hamiltonian of a ferromagnet in the external magnetic field:
\begin{equation}\label{ham0}
{\cal H}=\sum_{\langle i,j\rangle}[{\cal J}_zs_i^zs_j^z+{\cal J}_ys_i^ys_j^y]+h\sum_is_i^z + {\cal C}
\end{equation}
where $\langle i,j\rangle$ labels now nearest-neighbor pseudospins, arranged in a triangular lattice formed by the stars, $\bf s$ is spin-$\frac12$ operator, ${\cal C}=-5.268J_1{\cal N}$, ${\cal J}_z=4a_3=-0.007J_1$, ${\cal J}_y=4a_4=-0.001J_1$ and $h=-6a_2=-0.092J_1$. Here ${\cal N}=N/12$ is the number of stars in the lattice. Factor 6 appears in the expression for $h$ because each star interacts with 6 neighbors. One can see that the magnetic field in the effective Hamiltonian Eq.~(\ref{ham0}) is much more larger than the exchange. So in this approximation stars behave like almost free spins in the external magnetic field and the ground state of KAF has long-range singlet order which settles on the triangular star lattice and is formed by stars in $\Psi_-$ states.

{\it $V^3$ corrections}.---The field remains the largest term of the effective Hamiltonian and the KAF ground state has the same long-range order if one takes into account $V^3$ terms in the perturbation series. For two-stars coupling $V^3$ corrections have the form given by Eq.~(\ref{siecle}). Terms $V^3$ imply analysis of the three-stars interaction as well. Nonzero contributions from them one gets only for the configuration presented in Fig.~\ref{clusters}(b). The secular matrix for three stars is of the size $8\times8$. We have calculated $V^3$ corrections with a very high precision using the integral representation similar to that Eq.~(\ref{s2}) for the second term in Eq.~(\ref{siecle}). All the operator exponents were expanded up to the power 150. As a result the low-energy properties of the KAF are described by the effective Hamiltonian:
\begin{equation}\label{ham}
    {\cal H}=\sum_{\langle i,j\rangle}\left[{\cal J}_zs_i^zs_j^z+{\cal J}_xs_i^xs_j^x+{\cal J}_ys_i^ys_j^y\right]+h\sum_is_i^z + {\cal C},
\end{equation}
where all parameters are given in the Table~\ref{parameters}. It describes two-stars coupling. We omit the three pseudospins terms in Eq.~(\ref{ham}) which have the form $s^z_is^z_js^z_k$ and $s^z_is^y_js^y_k$ and describe three-stars interaction. The corresponding coefficients are of the order of $10^{-3}J_1$ and $10^{-4}J_1$, respectively, and are negligible in comparison with those of the presented terms. It should be stressed that within our precision Hamiltonian Eq.~(\ref{ham}) is an exact mapping of the original Heisenberg model in the low-energy sector (excitation energy $\omega\sim\max\{{\cal J}_z,{\cal J}_y,{\cal J}_x,h\}\ll J_1$).

As one can conclude from the Table~\ref{parameters} studying $V^3$ corrections the common shift given by them remains much larger than the levels splitting in both cases of two- and three-stars coupling. Meanwhile the values of $V^3$ perturbation terms are approximately two times smaller than those of $V^2$ ones. So the change in the effective Hamiltonian from $V^3$ terms is significant and analysis of the perturbation series can not be finished at this point for correct establishing of the effective Hamiltonian. Unfortunately such a work requires great computer capacity we do not dispose now. We have to restrict ourself with this precision here.

One can judge about the applicability of the perturbation series from the following values of the ground state energy of two interacting stars shown in Fig.~\ref{twostars} obtained numerically and using first two orders of perturbation theory. Ground state energy of two non-interacting stars is $-9J_1$. That of two interacting ones which we have calculated numerically by power method \cite{parlett} is $-9.62J_1$.  Whereas the ground state energy obtained using 
Table~\ref{parameters} is $-9.42J_1$ (contribution from $V^2$ and $V^3$ terms are $-0.27J_1$ and $-0.15J_1$, respectively).

{\it Effective Hamiltonian structure}.---Despite the perturbation theory works quite badly in the star model and many perturbation terms are to be taken into account, we can proof now that Eq.~(\ref{ham}) is the most general form of the effective Hamiltonian assuming that $n$-pseudospins couplings with $n>2$ are small as it was in the case of $n=3$ discussed above. Let us consider possible terms of the form $s^z_is^+_j$, $s^z_is^-_j$, $s^+_i$ and $s^-_i$. In these cases numbers of functions $\Psi_+$ and $\Psi_-$ to the right hand of the corresponding matrix elements differ from those to the left hand by unit. As it was also pointed out above, \kagome lattice contains lines of symmetry reflections and star Hamiltonian and inter-star interaction are invariant with respect to  these reflections. As $\Psi_+$ are invariant and $\Psi_-$ changes the sign under these transformations, the matrix elements are equal to themselves with the opposite sign and so must be zero. Other possible term $s^x_is^y_j=-(i/4)(s^+_is^+_j - s^-_is^-_j + s^-_is^+_j - s^+_is^-_j)$ can not appear in the effective Hamiltonian because the corresponding matrix elements should be imaginary.

{\it Ground state}.---As is clear from the Table~\ref{parameters}, ${\cal J}_x$ and $h$ are the largest parameters of the Hamiltonian Eq.~(\ref{ham}) in our approximation. So KAF behaves like Ising antiferromagnet in perpendicular magnetic field. In this case the classical value of the field at which spin-flip occurs is $h_{s-f}={\cal J}_x$ which is approximately 2.6 times smaller than $h$. So the ground state should remain ordered with all the stars in $\Psi_-$ state.

The ground state energy and that of the upper edge of the singlet band calculated using Table~\ref{parameters} are $(-4.5J_1+\Delta{\cal C}+h/2+3{\cal J}_z/4){\cal N}=-0.452J_1N$ and $(-4.5J_1+\Delta{\cal C}-h/2+3{\cal J}_z/4){\cal N}=-0.437J_1N$, respectively. Corrections from ${\cal J}_x$ to these values in the first non-zero order of perturbation theory are given by $(3/16){\cal N}{\cal J}_x^2/h$ and are negligible. At the same time the ground state energy of the largest cluster with $N=36$ considered before numerically was $-0.438J_1N$ \cite{wald}. So we believe that clusters used in the previous studies were too small to reflect the Heisenberg KAF low-energy sector at $J_2=0$ properly.

{\it Interaction $J_2$}.---We show now that in spite of its smallness the next-nearest neighbor interaction can play an important role for low-energy properties. We have calculated $J_2$ corrections to the parameters of the effective Hamiltonian Eq.~(\ref{ham}) for the first and for the second terms in Eq.~(\ref{siecle}) only. There are 12 intrinsic $J_2$ interactions in each star which splits the doubly degenerated ground state and, as is seen from the Table~\ref{parameters}, gives a contribution to the magnetic field $h$ and to the constant $\cal C$.

As is clear from Fig.~\ref{twostars} the two-stars coupling is given now by the operator $\tilde V=J_2({\bf S}^{(1)}_1{\bf S}^{(2)}_2+{\bf S}^{(1)}_2{\bf S}^{(2)}_1+{\bf S}^{(1)}_2{\bf S}^{(2)}_3+{\bf S}^{(1)}_3{\bf S}^{(2)}_2)$. Corrections proportional to $J_2$ were calculated by the same way as above and are presented in the
Table~\ref{parameters} too. It is seen that the contribution of next-nearest interactions to the magnetic field becomes significant if $|J_2|\sim 0.1J_1$. If $J_2<0$ (ferromagnet interaction) they can even change the sign of $h$.

Effect of next-nearest ferromagnetic coupling for KAF properties was previously studied in Ref.~\cite{lecheminant} numerically on finite clusters with $N\le 27$ in the wide range of the values of $J_2$. It was shown there that at $|J_2|/J_1\sim 1$ the ground state has $\sqrt{3}\times\sqrt{3}$ magnetic structure. At $|J_2|/J_1\ll 1$ the ground state was found to be disorder and there is a band of singlet excitations inside the triplet gap. As we demonstrated above in our approach this band is a result of star ground state degeneracy.

{\it $T^2$ specific heat behavior}.---There were many speculations on the low-temperature dependence of the KAF specific heat $C\propto T^2$ observed experimentally for $S=3/2$ (see Refs.~\cite{sind,zhit} and references therein). As we obtained above, low-$T$ properties are described by effective Hamiltonian Eq.~(\ref{ham}) of a magnet which has spectrum of the form $\epsilon_{\bf q}=\sqrt{(cq)^2+\Delta'^2}$ at $q\ll1$ and can be in ordered or disordered phases depending on particular values of the Hamiltonian parameters. Small $\Delta'$ here signifies the proximity to the quantum critical point in which $C\propto T^2$. Such a situation arises in the singlet dynamics of the model of interactive plaquets \cite{zhit} discussed above. We do not present here the corresponding analysis because parameters of the effective Hamiltonian could be changed in the further orders of the perturbation theory.

{\it Experimental verification}.---In both cases of ordered ground state and disordered one the approach presented in this paper can be checked by inelastic neutron scattering: corresponding intensity for the singlet-triplet transitions should have periodicity in the reciprocal space corresponding to the star lattice. This picture is similar to observed one in the dimerized spin-Pairls compound $\rm CuGeO_3$ \cite{germ}. In this case inelastic magnetic scattering has a periodicity which corresponds to the dimerized lattice.

{\it Cases of $S>1/2$}.---It should be noted that our consideration of $S=1/2$ KAF can not be expanded directly on the cases of larger spins. Although functions presented in Fig.~\ref{grounds}, where a bold line denotes the singlet state of the corresponding two spins, remain eigenfunctions of the Hamiltonian for $S>1/2$, we have found numerically that they are not ground states of the star with $S=1$ and $S=3/2$. All details of calculations are presented in Appendix~\ref{val_det}. So an other approach should be proposed for KAFs with $S>1/2$.

\section{Conclusion}
\label{con}

In this paper we present a model for the low-energy physics of spin-$\frac12$ \kagome Heisenberg antiferromagnets (KAFs). The spin lattice can be represented as a set of stars which are arranged in a triangular lattice and contain 12 spins (see Fig.~\ref{lattice}). Each star has two degenerate singlet ground states with different symmetry which can be described in terms of pseudospin. It is shown that interaction between the stars leads to the band of singlet excitations which determines the KAF low-energy properties. The low-energy dynamics is described Hamiltonian of a spin-$\frac12$ magnet in the external magnetic field given by Eq.~(\ref{ham}). The Hamiltonian parameters are calculated in the first three orders of perturbation theory and are summarized in the Table~\ref{parameters}. Within our precision KAF has an ordered singlet ground state with all the stars in the state given by Eq.~(\ref{f2}). The ground state energy is lower than that calculated in previous finite clusters studies. We show that our model can not be expanded directly on KAFs with $S>1/2$.

The approach discussed in this paper can be verified experimentally on inelastic neutron scattering: the corresponding intensities for singlet-triplet transitions should have periodicity in the reciprocal space corresponding to the star lattice.

We are grateful to A.G.\ Yashenkin for interesting discussions. This work was supported by Russian State Program "Collective and Quantum Effects in Condensed Matter", the Russian Foundation for Basic Research (Grants No.\ 03-02-17340, 00-15-96814) and Russian State Program "Quantum Macrophysics".

\appendix

\section{Irreducible representations of the group $C_{6v}$}
\label{ir}

The symmetry group $C_{6v}$ contains six rotations $C^k$ on the angles $2\pi k/6$ ($k=0,1,..,5$) and six reflections which can be written as $C^ku_1$, where $u_1$ is operator of a reflection. One-dimensional irreducible representations can be presented as follows \cite{petrashen,landau}:
\begin{align}
C^k&\sim 1, & u_1&\sim 1;\label{forpluse}\\
C^k&\sim (-1)^k, & u_1&\sim 1;\\
C^k&\sim 1, & u_1&\sim -1;\label{forminus}\\
C^k&\sim (-1)^k, & u_1&\sim -1.
\end{align}
For two-dimensional ones we have \cite{petrashen,landau}
\begin{align}
\label{two}
C^k & \sim
\left(
\begin{array}{cc}
e^{i\frac{2\pi l}{6}k} & 0 \\
0 & e^{-i\frac{2\pi l}{6}k}
\end{array}
\right),
 & u_1 & \sim
\left(
\begin{array}{cc}
0 & 1\\
1 & 0
\end{array}
\right),
\end{align}
where two inequivalent representations are given by $l=1$ and $l=2$.

\section{Star with $S=1$ and $S=3/2$}
\label{val_det}

\begin{figure}
  \centering
  \includegraphics{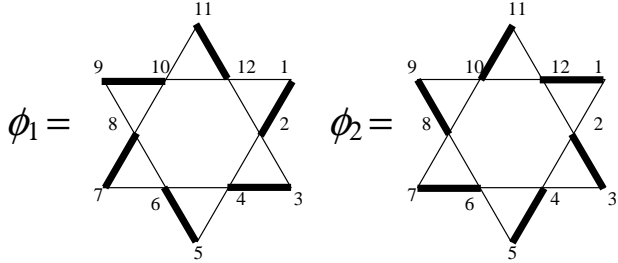}
  \caption
{
Schematic representation of a star's two singlet ground state wave functions $\phi_1$ and $\phi_2$.  A bold
line denotes the singlet state of two neighboring spins, i.e. $({\mid\uparrow\rangle}_i
{\mid\downarrow\rangle}_j - {\mid\downarrow\rangle}_i{\mid\uparrow\rangle}_j)/\sqrt{2}$.
\label{grounds}
}
\end{figure}

In this Appendix we present details of numerical calculations shown that functions presented in Fig.~\ref{grounds}, where a bold line denotes the singlet state of the corresponding two spins, are not ground states of the star with $S=1$ and $S=3/2$ as it was for $S=1/2$.

A simple numerical method for determination of the maximum by module eigenvalue of a Hermitian operator $H$ (power method \cite{parlett}) was used. It is based on the following statement. Let us consider a state of the system $f=\sum_ic_i\psi_i$, where the sum may not include all the $H$ eigenfunctions. The maximum by module eigenvalue $E_{extr}$ for given $f$ is determined by the expression
\begin{equation}
\label{simple}
\lim_{n\to\infty}\frac{\langle f|H^{n+1}|f\rangle}{\langle f|H^n|f\rangle}=E_{extr}.
\end{equation}
It becomes evident if one notes that $\langle f|H^n|f\rangle=\sum_i|c_i|^2E_i^n$.

The Eq.~(\ref{simple}) can be used in numerical calculations in the following way. The corresponding expression is calculated for $n=1,2,..,n_{max}$. So one can control the convergence comparing results with different $n$. Studying a full set of vectors $f$ and taking large enough $n_{max}$ to match the necessary precision one can find the largest by module eigenvalue of $H$.

In the case of the star the maximum eigenvalue of the Hamiltonian is $E_{max}=18S^2J_1$ (this energy has the state in which all the spins are along the same direction) and the energy of singlet states shown in Fig.~\ref{grounds} is $E_{ss}=-6S(S+1)J_1$. As $E_{max}>|E_{ss}|$ for $S>1/2$, we have to take $H={\cal H}_0-WI$ to investigate the lower ${\cal H}_0$ levels, where ${\cal H}_0$ is the star Hamiltonian given by Eq.~(\ref{h}), $I$ is an identical matrix and $W=(E_{max}+E_{ss})/2+J_1$. Thus eigenvalues of $H$ are shifted down according to those of ${\cal H}_0$ by the same value $W$ so as the largest by module $H$ eigenvalue to be equal to the ${\cal H}_0$ ground state energy minus $W$.

We have not found the ground state energy for the star with $S=1$ and $S=3/2$ by this method because a full set of vectors $f$ should be examined for that. This operation demands much computer time. But we have obtained studying a number of vectors $f$ that there are states lower by the energy at least by the value $1.8J_1$ than those discussed above. The method gave $E_{extr}$ with the prescribed precision to the second decimal place at $n_{max}=100\div300$ depending on $f$ and $S$.

\bibliography{kagome_prb}

\begin{table*}
\caption{
\label{parameters}
Contributions to parameters of the effective Hamiltonian Eq.~(\ref{ham}) from terms $V^1$, $V^2$ and $V^3$ of the perturbation expansion. Interaction $J_2$ has been taken into account in $V^1$ and $V^2$ terms only. Here ${\cal N}$ is the number of stars in the lattice.
}
\begin{ruledtabular}
\begin{tabular}{cccccc}
	&\multicolumn{1}{c}{$V^1$}&\multicolumn{1}{c}{$V^2$}&\multicolumn{2}{c}{$V^3$\footnotemark[1]} 				 		&\multicolumn{1}{c}{Totals} \\
		& & & two-stars & three-stars \\
		\hline
		${\cal J}_z$ & 0 		& $-0.007J_1+0.002J_2$ & $-0.013J_1$ & $0.010J_1$ & $-0.010J_1+0.002J_2$ \\
		${\cal J}_y$ & 0 		& $-0.001J_1+0.007J_2$	& $-0.001J_1$ & $0.001J_1$ & $-0.001J_1+0.007J_2$ \\
		${\cal J}_x$ & 0 		& 0					 & $0.067J_1$ & 0 & $0.067J_1$ \\
		$h$          & $-0.563J_2$ & $-0.092J_1-0.218J_2$ & $-0.161J_1$ & $0.080J_1$ & $-0.173J_1-0.781J_2$ \\
		$\Delta\cal C$\footnotemark[2]     & $-0.009J_2{\cal N}$ 		& $-0.768J_1{\cal N}+1.530J_2{\cal N}$ & $-0.361J_1{\cal N}$ & $0.304J_1{\cal N}$ & $-0.825J_1{\cal N}+1.521J_2{\cal N}$ \\
\end{tabular}
\end{ruledtabular}
\footnotetext[1]{This term implies two stars coupling shown in Fig.~\ref{twostars} and three stars interaction in the configuration presented in Fig.~\ref{clusters}(b).}
\footnotetext[2]{Correction to the value ${\cal C}_0=-4.5J_1{\cal N}$ for non-interactive stars.}
\end{table*}

\end{document}